# THERMOPOWER IN OVERDOPED REGION OF CUPRATES

Zorica KONSTANTINOVIĆ[1], Guénaëlle LE BRAS-JASMIN[1], Dorothée COLSON[1], Anne FORGET[1], Frédéric JEAN[2,3], Gaston COLLIN[3], Claude AYACHE[1]

[1]SPEC, DSM/DRECAM, CEA-Saclay, 91191 Gif sur Yvette, France
[2]LEMHE, Université Paris-SUD, Bât 415, 91405 Orsay, France
[3]LLB, DSM/DRECAM, CEA-CNRS, CEA-Saclay, 91191 Gif sur Yvette, France

The thermoelectric power S(T) of single-layer $Bi_2Sr_2CuO_{6+\delta}$ is studied as a function of oxygen doping in the strongly overdoped region of the phase diagram (T, $\delta$). As other physical properties in this region, diffusion thermopower $S_{diff}(T)$ also shows an important deviation from conventional Fermi liquid behaviour. This departure from T-linear S(T) dependence together with the results of susceptibility on the same samples suggest that the origin of the observed non-metallic behaviour is the existence of a singularity in the density of states near the Fermi level. The doping and temperature dependence of themopower is compared with a tight-binding band model.

1. INTRODUCTION

Doping dependence of thermoelectric power (TEP) at 290K, $S_{290}$, given by the universal Obertelli-Cooper-Tallon law[1] (OCT), is currently used as a measure of the hole concentration per $CuO_2$ plane, p, which plays a key role in the superconductivity. This robust $S_{290}$ vs p relation seems to be valid even in case of $YBa_2Cu_3O_{7-\delta}$ where the charge is distributed between CuO chains and $CuO_2$ planes. The only compound where this universal OCT law fails is the untypical $La_{2-x}Sr_xCuO_4$ (LSCO) superconductor[2]. However, quite recently, a renewed interest has been raised by the observation of a departure from OCT trend in the case of $Bi_2Sr_{2-z}La_zCuO_{6+\delta}$ (BSLCO)[3].

The strongly two-dimensional character of cuprates has suggested the interpretation of their physical properties in the frame of a tight-binding band model of the $CuO_2$ square lattice involving a van Hove singularity (VHS) in the density of states (DOS)[4]. This description is supported by recent photoemission measurements, that have indicated in several overdoped cuprates the existence of a saddle point[5,6,7] at $(0,\pi)$ and $(\pi,0)$ giving rise to the VHS. The general doping TEP tendency has been anticipated on the basis of this model[8] with a divergence of TEP occurring when the VHS crosses the Fermi level. Although still under discussion, the temperature behaviour of TEP supports a model based on two different drag and diffusion additive contributions[9,10]. In the overdoped region, S(T) on BSLCO compounds[11] and $Bi_2Sr_2CuO_{6+\delta}$ (BSCO)[12] shows deviations from usually reported T-linear behaviour as it is suggested by recent theoretical studies[13,14].

The present work concentrates on the doping evolution of the thermoelectric power on $Bi_2Sr_2CuO_{6+\delta}$ in the overdoped region. The strongly overdoped samples show an important deviation from T-linear metallic behaviour in the diffusion thermopower. The experimental results are analysed by using the one tight-binding band model. This model predict an anomaly in the doping dependence of $S_{diff}$ and a departure from the OCT law in the proximity of the van Hove singularity.

2. EXPERIMENTAL DETAILS

Polycrystalline $Bi_2Sr_2CuO_{6+\delta}$ samples were prepared by the classical solid reaction method[15]. All samples have the same cationic composition and their phase purity is controlled by X-ray diffraction. The non-substituted BSCO compound is intrinsically strongly overdoped (0.20<p<0.28). Hole concentration can be adjusted in a reversible way through oxygen excess $\delta$ which has been determined by thermogravimetric techniques[15]. The absence of CuO chains in these compounds allows to associate the observed properties directly to the $CuO_2$ planes. The thermoelectric measurements were performed by a conventional steady flow technique. Temperature and voltage gradients were simultaneously measured by T-type thermocouples. The critical temperature is determined from the onset of dc magnetisation measurements.

3. RESULTS

In Fig.1 we show the doping dependence of the room temperature thermoelectric power, $S_{290K}$ (connected close circles). The most overdoped sample $Bi_2Sr_2CuO_{6.18}$ is located at the limit of the superconducting region with $T_c$<1.5 K (p~0.28), while the less overdoped sample $Bi_2Sr_2CuO_{6.09}$ is close to the optimal doping (p~0.20). From previous analysis[12], the hole number p is found to be related to oxygen excess $\delta$ (indicated at top axes) by p=0.125+0.88$\delta$ (within an error of 10 %). As drag contribution appears for less overdoped samples ($\delta$<0.175), the extracted diffusion values[12] at 290 K are also shown in the same figure (open circles). As contribution of drag increases with decreasing doping[10,12], the diffusion part increases slower than the total thermopower. Even for the less overdoped sample, large negative value has been found for diffusion TEP.

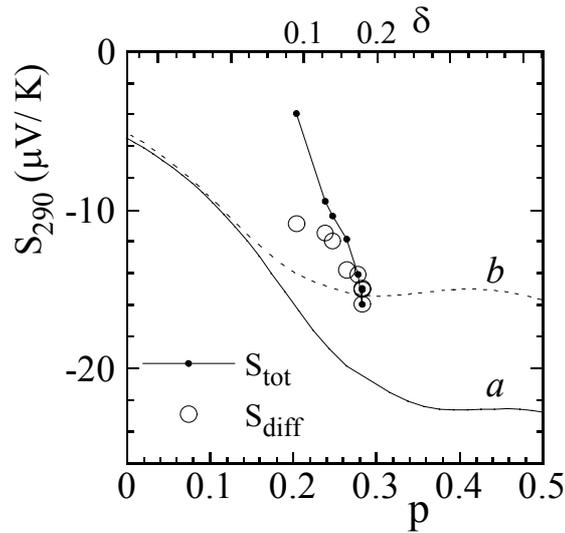

FIGURE 1
The doping dependence of the total (connected close circles) and diffusion (open circles) thermopower at 290 K. The δ values are indicated in top axis. The $S_{diff}$ calculated from tight-binding band model with t=0.35 eV and t'=0.11 eV (solid line, *a*) and t'=0.09 eV (dashed line, *b*) is also shown.

To analyse the temperature variation of TEP, we choose the most overdoped sample $Bi_2Sr_2CuO_{6.18}$, where the drag contribution is absent (Fig. 2). This diffusion thermopower $S_{diff}(T)$ clearly shows a deviation from the usually reported T-linear metallic behaviour and it can be approximately described by a phenomenological law of the form[12] $S_{diff}=T/(A+BT)$.

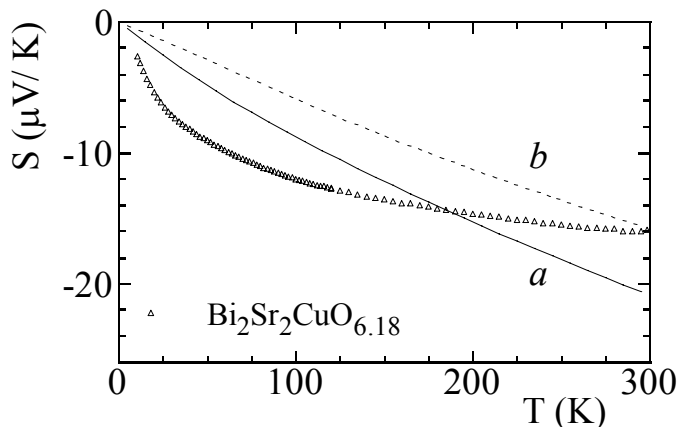

FIGURE 2
S(T) of $Bi_2Sr_2CuO_{6.18}$ ceramics (triangle). The T-dependence of thermopower is compared with calculated S(T) using band model for p~0.28 with t=0.35 eV (*a*, t'=0.11 eV and *b*, t'=0.09 eV) .

This behaviour of the diffusion thermopower indicates the presence of a narrow-band near the Fermi level, which is consistent with the doping and temperature variation of the susceptibility on the same samples[16]. The presence of this narrow-band is supported by recent photoemission measurements[5-7], indicating the van Hove singularity in the DOS close to the Fermi level.

## 4. DISCUSSION

We compare our results with the simplest model of the conduction band of the cuprates which can take into account the photoemission results[5-7] and the existence of a van Hove singularity at a realistic value of the band filling. The energy dispersion of a square-lattice in the tight-binding model[8] with nearest- and next-nearest-neighbour hopping t and t', respectively, is given by $\varepsilon(\mathbf{k})=-2t(\cos(k_xa)+\cos(k_ya))+4t'\cos(k_xa)\cos(k_ya)$, where a is the lattice parameter. The band width W is determined by the value of nearest-neighbour interaction W=8t, while the parameter t' determines the position of the VHS ($E_{VHS}$=-4t'). The ratio between t/t' is related to the hole number where the Fermi level crosses VHS, corresponding to the change from hole-like to electron-like Fermi surface.

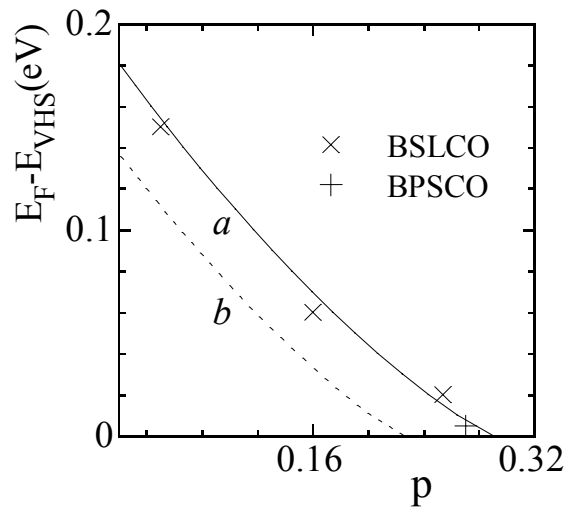

FIGURE 3
The doping dependence of $E_F$-$E_{VHS}$, estimated from ARPES on single layer BSLCO[6] (pluses) and BPSCO[7] (cross) single crystals, is fitted using tight-binding band model with t=0.35 eV and t'=0.11 eV (*a*, solid line). Calculation with t=0.35 and t'=0.09 eV is also shown (*b*, dashed line).

In Fig.3, we fit the energy shift of the VHS from the Fermi level, $E_F-E_{VHS}$, estimated from photoemission measurements on BSLCO[6] (crosses) and BPSCO[7] (plus) single crystals with the described band model and by assuming that t and t' do not change with doping. The good agreement with ARPES data is obtained for values of the nearest- and next-nearest-neighboured interactions of t=0.35 eV and t'=0.11 eV (solid line, *a*). These values lead to the topological change of the Fermi surface ($E_F=E_{VHS}$) at hole number of p~0.29. For comparison, energy shift $E_F-E_{VHS}$ is also shown for parameters t=0.35 eV and t'=0.09 eV (dashed line, *b*) which correspond to a topological change at p~0.22, as seen in the LSCO system. Now, these sets of values (t,t'), *a* and *b*, will be used to calculate the doping and temperature variation of TEP.

The diffusion thermopower is given by the standard expression[17], $S_{diff} = \frac{1}{eT\sigma} \int (\varepsilon-\mu)\sigma(\varepsilon)\frac{\partial f_0}{\partial \varepsilon} d\varepsilon$, where $\sigma(\varepsilon)$ is a partial conductivity function at energy $\varepsilon$, $\mu$ is the chemical potential, e is the electronic charge and $f_0$ is the Fermi-Dirac function. In our calculation the relaxation time $\tau(\mathbf{k},\varepsilon)$ is assumed to be constant. The solid and the dashed line in Fig.1 show that the calculated doping variation of $S_{diff}$ at 290 K is in qualitatively good agreement with the diffusion TEP determined from experimental values. Notice that instead of a divergence[8] when $E_F$ lies on $E_{VHS}$, room temperature $S_{diff}$ presents an anomaly in the proximity of VHS. This prediction gives a limit of validity to the OCT law[1]. Thus, the linear variation of S with p is no longer applicable in the strongly overdoped region when the VHS is close to the Fermi level.

At the same time, the temperature dependence of the calculated TEP shows a clear deviation from T-linear behaviour, but it is less pronounced than our experimental results. Further improvements are then needed as, for example, taking into account the energy dependence of the relaxation time[18] or the incommensurate superstructure in the BiO layer leading to an umklapp scattering.

5. SUMMARY

We have presented the doping evolution of the thermoelectric power S(T) of $Bi_2Sr_2CuO_{6+\delta}$ ceramics in the strongly overdoped region. The diffusion thermopower $S_{diff}(T)$ shows a deviation from usually reported T-linear variation, which indicates the presence of a narrow band near the Fermi level. The results are analysed in terms of a tight-binding band model leading to a qualitatively good description of both the doping and the temperature dependence of the TEP.

ACKNOWLEDGEMENT

We are grateful to C. Chaleil and L. Le Pape for their technical support. We thank M. Norman, C. Pepin, J. Bouvier, J. Bok, M. Roger and J.P. Carton for stimulating discussions.